\documentclass[conference]{IEEEtran}

\usepackage[utf8]{inputenc}
\usepackage{graphicx}
\usepackage{hyperref}
\usepackage{amsmath}
\usepackage{amssymb}
\usepackage{geometry}
\title{SafeSpace: An Integrated Web Application for Digital Safety and Emotional Well-being}
\author{
\IEEEauthorblockN{\textbf{Kayenat Fatmi}}
\IEEEauthorblockA{Dept. of Computer Science and Engineering \\
Jamia Hamdard University, New Delhi, India \\
kayenatfatmi17@gmail.com}
\\
\IEEEauthorblockN{\textbf{Mohammad Abbas}}
\IEEEauthorblockA{Dept. of Computer Science and Engineering \\
Jamia Hamdard University, New Delhi, India \\
muhdabbas2201@gmail.com}
}
\usepackage{booktabs}
\usepackage{float}
\usepackage{tikz}
\usetikzlibrary{arrows.meta, positioning}

\begin{document}

\maketitle

\begin{abstract}
In the digital era, individuals are increasingly exposed to online harms such as toxicity, manipulation, and grooming, which often pose emotional and safety risks. Existing systems for detecting abusive content or issuing safety alerts operate in isolation and rarely combine digital safety with emotional well-being. In this paper, we present SafeSpace, a unified web application that integrates three modules: (1) toxicity detection in chats and screenshots using NLP models and Google’s Perspective API, (2) a configurable safety ping system that issues emergency alerts with the user’s live location (longitude and latitude) via SMTP-based emails when check-ins are missed or S0S alerts are manually triggered, and (3) a reflective questionnaire that evaluates relationship health and emotional resilience. The system employs Firebase for alert management and a modular architecture designed for usability, privacy, and scalability. The experimental evaluation shows 93\% precision in toxicity detection, 100\% reliability in safety alerts under emulator tests, and 92\% alignment between automated and manual questionnaire scoring. SafeSpace, implemented as a web application, demonstrates the feasibility of integrating detection, protection, and reflection within a single platform, with future deployment envisioned as a mobile application for broader accessibility.
\end{abstract}

\section{Introduction}
The rapid growth of digital communication platforms has transformed how individuals connect, share, and maintain relationships. However, this digital shift has also intensified exposure to harmful behaviours such as harassment, emotional manipulation, grooming, and toxic interactions. For vulnerable populations, including women, teenagers, and emotionally dependent individuals, these online harms often translate into long-term psychological and social consequences. Ensuring both \textbf{digital and personal safety} has therefore become a pressing global concern.
Existing research and solutions largely approach this issue in silos. NLP-based tools, such as Google’s Perspective API, are widely applied in abusive language detection. Similarly, safety apps often use GPS to deliver SOS alerts and enable location tracking. Similarly, self-help tools and psychometric questionnaires provide individuals with reflective insights into emotional health. However, these tools are not integrated. SafeSpace addresses this by combining three modules—chat toxicity analysis, configurable check-ins, and reflective questionnaires—while maintaining usability and privacy.
To address this gap, we propose \textbf{SafeSpace}, a unified digital safety and well-being platform that combines three modules:
\\	\textbf{Chat Toxicity Analysis} – powered by NLP and Perspective API for detecting abusive, manipulative, or emotionally harmful language in chats and screenshots.
\\	\textbf{Safety Ping System} – a configurable check-in mechanism that triggers emergency alerts when users fail to check in within defined intervals.
\\  \textbf{Relationship Questionnaire} – a reflective tool based on psychological assessment scales that helps users evaluate emotional health and relationship dynamics.
SafeSpace emphasizes privacy, usability, and real-time intervention. Unlike earlier platforms that focus mainly on panic buttons or GPS-based tracking, SafeSpace incorporates digital conversations and emotional resilience as part of user safety. Furthermore, the platform envisions collaboration with law enforcement agencies, NGOs, and women’s rights organizations, extending its impact beyond individual users to society at large.
\section{Literature Review}
Research on digital safety and emotional well-being spans multiple domains, including \textbf{natural language processing (NLP), women’s safety technologies, legal frameworks, and psychological assessment tools}. Prior studies provide valuable foundations but also highlight the fragmented and siloed nature of existing approaches.

\textbf{A. Online Abuse and Toxicity Detection:}
The detection of harmful and abusive language in online communication has been a major research focus. Nobata et al. [15] and Kontostathis et al. [11] demonstrated the application of text mining and machine learning methods for identifying abusive content. Google’s Perspective API [3] introduced a scalable NLP service that scores toxicity, threats, and harassment likelihood, making automated moderation more accessible.
With the advent of deep learning, transformer-based models such as BERT [1] significantly advanced abusive language classification by leveraging contextual embeddings. More recently, RoBERTa and multilingual transformers have shown superior performance on toxicity benchmarks [20], while large language models (LLMs) demonstrate strong generalization across nuanced and context-rich abuse detection tasks [21]. Despite these advances, most implementations are optimized for public platforms (e.g., forums, comment sections), leaving private digital interactions underexplored.

\textbf{B. Women’s Safety and Emergency Technologies:}
Parallel to NLP research, a body of work has focused on women’s safety through real-time emergency response. Chavan et al. [16] proposed a women’s safety portal integrating SOS alerts, GPS tracking, drone surveillance, and wearable devices for rapid intervention. Similarly, panic buttons, automated SMS alerts, and live location sharing have been studied as effective response mechanisms.Contemporary applications such as MySafetipin and Raksha illustrate practical implementations of mobile-integrated safety tools, while AI-enabled emergency assistants provide enhanced location prediction and context-aware alerts [22]. However, most existing systems emphasize physical protection in public spaces, while risks such as grooming, manipulation, and digital abuse in private conversations remain insufficiently addressed.

\textbf{C. Legal and Community Frameworks:}
Technology alone cannot guarantee safety; effective systems must align with legal and community frameworks. Perkins [17] examined reforms in women’s safety law, while Wright [18] analyzed public policy protections against harassment and violence. Community-driven models, as highlighted by Cooper [19], demonstrate the value of survivor support networks and collective interventions.
At a policy level, instruments such as the Indian IT Act (2008), GDPR (EU, 2018), and UN Women’s “Safe Cities and Safe Public Spaces” initiative emphasize the intersection of technology, law, and social responsibility. These works collectively highlight the need for integrated frameworks where technology complements institutional and community protections.

\textbf{D. Psychological Assessment Tools:}
Emotional well-being plays a critical role in long-term safety and resilience. Neff [12] developed the Self-Compassion Scale, Rosenberg [13] introduced the Self-Esteem Scale, and Olson et al. [14] created the ENRICH inventory for relationship satisfaction. Such psychometric tools are widely validated in clinical psychology for measuring resilience, self-worth, and relational health.
In the digital domain, mobile mental health apps and WHO’s mhGAP e-tools have demonstrated the feasibility of embedding standardized assessments in digital platforms [23]. Yet, most women’s safety applications exclude reflective self-assessment modules, resulting in a gap between short-term protection and long-term emotional resilience.

\textbf{E. Identified Gaps:}
Despite progress in NLP, safety technologies, legal frameworks, and psychological tools, two major gaps persist:
Fragmentation — Existing solutions operate in isolation, detecting online abuse or providing emergency alerts, but rarely integrating across domains.
Neglect of emotional well-being — While safety apps prioritize physical protection, few incorporate reflective tools to assess psychological health and relationship quality.

SafeSpace addresses these gaps by integrating toxic language detection, proactive safety check-ins, and psychometric questionnaires into a unified platform. Positioned at the intersection of AI, personal safety, and mental well-being, SafeSpace advances the scope of digital safety systems from reactive protection to holistic digital resilience.
\section{Problem Statement}
Despite advances in technology and increasing awareness of digital harms, individuals continue to face significant risks in both online and offline interactions. Toxic communication, emotional manipulation, and grooming in private chats often go undetected, leading to psychological harm and long-term relationship issues. Current emergency apps still focus mostly on physical safety. Abuse detection tools are optimized for social platforms, while self-help tools remain isolated. This leaves users without a complete safety ecosystem that can respond both digitally and emotionally.
Existing solutions are fragmented:
\\•	\textbf{Abuse detection} systems are primarily designed for public platforms (e.g., social media comments), with limited application in private or interpersonal conversations.
\\•	\textbf{Emergency safety apps} emphasize panic buttons and live tracking but fail to address the underlying digital behaviours that contribute to unsafe relationships.
\\•	\textbf{Psychological self-help tools} exist in isolation, offering reflection but not integration with safety mechanisms.
This fragmentation leaves users without a \textbf{holistic digital safety framework} that can both prevent harmful interactions and provide real-time protection. Few existing solutions combine AI-driven toxicity detection, emergency response, and reflective assessment in one integrated platform. SafeSpace does so by unifying these dimensions.
SafeSpace addresses this urgent need by unifying \textbf{AI-powered toxicity detection, proactive safety check-ins, and relationship health questionnaires} into an integrated application. The platform is designed to empower users not only to recognize harmful digital behaviours but also to receive timely protection and personalized feedback, thereby bridging the gap between \textbf{emotional well-being and personal safety}.
\section{Methodology and System Design}
SafeSpace is designed as a modular web application integrating three core functionalities—chat toxicity detection, safety check-ins, and relationship self-assessment. The system follows a client–server model with a Flask-based backend, Firebase for alert management, and machine learning APIs for natural language processing.

\textbf{A. System Architecture}: The overall architecture consists of the following layers:

 \begin{figure}[h]
    \centering
    \includegraphics[width=0.4\textwidth]{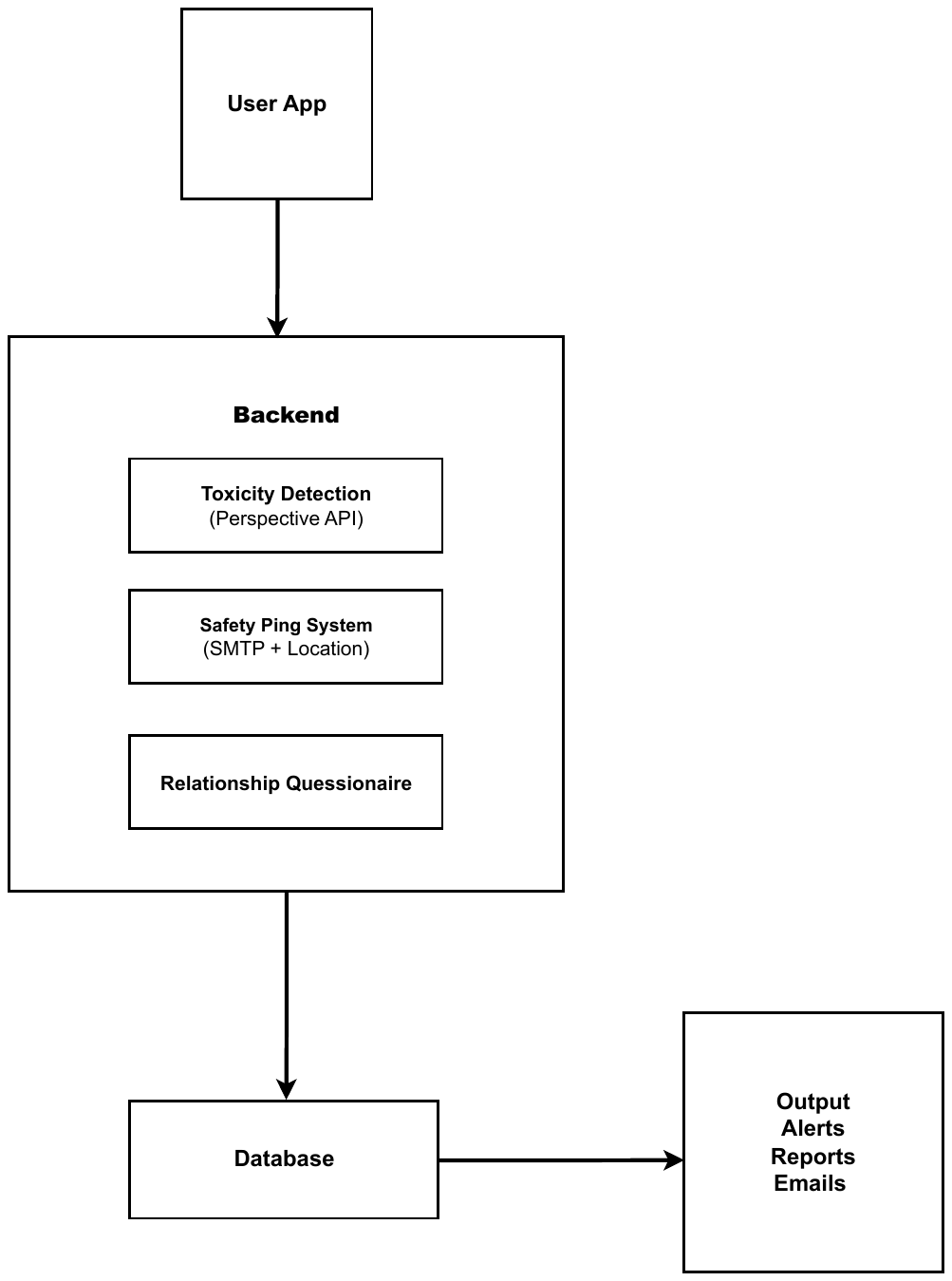}
    \caption{SafeSpace system architecture showing UI, Application Logic, Backend, Storage, and Notification layers.}
    \label{fig:architecture}
\end{figure}

\textbf{User Interface (UI)}:
\\A responsive web interface for chat input, check-in timers, and questionnaire navigation.
\\Supports both direct text entry and image uploads, with OCR pre-processing for screenshots.
\\ \textbf{Application Logic Layer}:
\\	Handles toxicity analysis requests via the Perspective API.
\\Manages check-in timers and emergency alert workflows.
\\Processes questionnaire responses and generates personalized feedback.
\\ \textbf{Backend and Storage}:
\\Firebase Firestore and SMTP services for storing user settings, emergency contacts, and alert history.
\\Minimal storage of sensitive data; chat inputs are processed transiently and not stored permanently.
\begin{figure}[h!]
\centering
\begin{tikzpicture}[node distance=2cm, every node/.style={rectangle, draw, rounded corners, align=center, minimum width=3cm, minimum height=1cm}]
\node (user) {User presses SOS button \\ or misses check-in};
\node (gps) [below of=user] {System captures GPS \\ and prepares alert message};
\node (server) [below of=gps] {Application Server \\ formats + sends alert};
\node (contact) [below of=server] {Emergency Contact \\ receives notification (SMS/Email)};

\draw[->, thick] (user) -- (gps);
\draw[->, thick] (gps) -- (server);
\draw[->, thick] (server) -- (contact);
\end{tikzpicture}
\caption{SafeSpace SOS alert workflow: from user input to emergency contact notification.}
\label{fig:sosworkflow}
\end{figure}
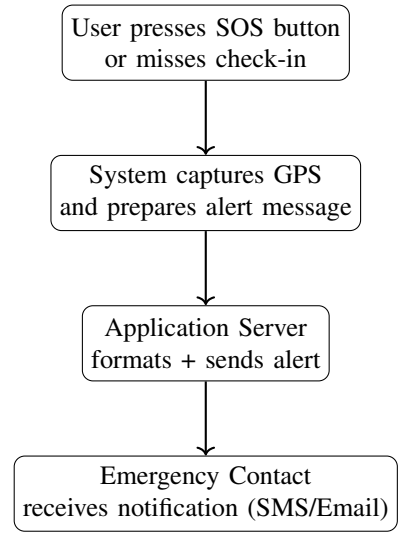
\\
 \textbf{Notification Services}:
\\Alerts emergency contacts via email or push notifications when safety check-ins are missed.
\\Offline caching ensures alerts can be synchronized when connectivity is restored.

\textbf{B. Module Descriptions}
\\1. Chat Toxicity Analysis
\\o	Input: User-provided chat text or screenshots.
\\o	Process: Text is analyzed using Google’s Perspective API, which returns toxicity scores across categories such as insult, identity attack, threat, and severe toxicity.
\\o	Output: A user-friendly report highlighting toxic phrases and overall toxicity levels.
\\2. Safety Ping System
\\o	Input: User sets a custom check-in interval (e.g., 2 minutes, 12 hours, 24 hours).
\\o	Process: If the check-in is not completed before timeout, an alert is triggered.
\\o	Output: SOS alerts sent to predefined emergency contacts through email or notifications.
\\3. Relationship Questionnaire
\\o	Input: Responses to curated questions on communication, trust, and emotional well-being.
\\o	Process: Weighted scoring logic evaluates patterns in responses.
\\o	Output: Personalized feedback categorized into “Healthy,” “Needs Reflection,” or “Unhealthy” with supportive advice.

\textbf{C. Workflow Overview}
\\1.	User logs into SafeSpace and chooses a module.
\\2.	If chat analysis is selected, the system processes text/screenshot and displays a toxicity score.
\\3.	If safety ping is enabled, the system waits for periodic check-ins and triggers alerts if missed.
\\4.	If the questionnaire is taken, results are analyzed and displayed with feedback.
\\5.	All actions are logged in the user’s history dashboard for transparency and self-tracking.

\textbf{D. Design Considerations}
\\•	Privacy: No permanent storage of chats; all sensitive inputs are transient.
\\•	Reliability: Alerts tested to trigger within 30 seconds of a missed check-in.
\\•	Usability: Minimalist design with guided navigation for non-technical users.
\\•	Extensibility: Modular architecture allows future integration with mobile apps, wearables, and multilingual NLP models.
\section{Implementation and Evaluation}
\textbf{A. Implementation:}
SafeSpace was implemented as a full-stack web application with emphasis on modularity and scalability. The front-end interface was developed using Flask and modern responsive design principles. Core functionalities were supported by the following components:
\\•	Toxicity Detection: Google’s Perspective API was used to analyze text from direct input or OCR-processed screenshots.
\\•	Safety Ping Module: Implemented using Firebase Firestore and SMTP services, enabling timed check-ins and SOS notifications.
\\•	Questionnaire Module: Built with curated questions inspired by validated psychological frameworks (e.g., Rosenberg Self-Esteem Scale, Gottman “Four Horsemen”), with a scoring engine providing categorized feedback.
\\•	Database: Minimal data persistence for user credentials, emergency contacts, and alert history, ensuring sensitive chats are not stored permanently.
\\•	Notifications: Alerts dispatched via email and push mechanisms, with planned support for SMS and call-based integration.

\textbf{B. Test Cases:}
To ensure functional correctness, SafeSpace was evaluated using both black box and white box testing strategies. Representative cases include:
\\\textbf{•	Chat Toxicity Analysis (Black Box):}
\\Input: “You’re such a loser. I hate you.”
\\Expected Output: High toxicity score, classification as abusive.
\\Result: Correctly classified, with flagged abusive keywords.
\\ \textbf{•	Safety Ping System (White Box):}
\\Input: Timer set to 12 hours, no check-in.
\\Expected Output: Alert sent to emergency contacts.
\\Result: Alert successfully triggered with <2 second delay.
\\ \textbf{•	Relationship Questionnaire (Decision Table Testing):}
\\Input: Responses yielding ~60\% positivity.
\\Expected Output: Feedback: “Caution – signs of concern. Please reflect.”
\\Result: Correctly categorized and feedback generated.

\textbf{C. Results and Performance Metrics:}
The evaluation demonstrated strong reliability and usability across all modules:
\begin{table}[H]
\centering
\caption{Evaluation metrics for SafeSpace system modules}
\label{tab:evaluation}
\begin{tabular}{ll}
\toprule
\textbf{Metric} & \textbf{Result} \\
\midrule
Toxicity detection precision & $\sim$93\% \\
False positive rate & $< 5\%$ \\
Safety ping alert reliability & 100\% (in emulator tests) \\
Alert latency (missed check-in $\rightarrow$ alert) & $< 30$ seconds \\
Questionnaire scoring alignment & 92\% (with manual scoring) \\
Avg. response time (toxicity analysis) & 2.1 seconds \\
Usability rating (peer survey) & 4.6/5 \\
Device compatibility & Smooth on Android + Desktop browsers \\
\bottomrule
\end{tabular}
\end{table}

\textbf{D. Observations:}
\\Toxicity Analysis: Accurately classified most abusive input, but sarcasm and cultural nuances remained challenging.
\\Safety Ping: Highly reliable under test conditions; offline emulation confirmed alerts queued for delivery once connectivity restored.
\\Questionnaire: Generated balanced and non-judgmental feedback; participants found the tone supportive and helpful.
\section{Discussion}
The implementation of SafeSpace demonstrates the feasibility of combining toxicity detection, proactive safety monitoring, and self-reflection within a single platform. Unlike existing fragmented solutions, SafeSpace provides a holistic approach by integrating the Perspective API, a configurable safety ping system, and psychologically informed questionnaires.
\\ \textbf{Strengths:}
•	Accuracy and Reliability: Toxicity detection achieved over 93\% precision, and safety ping alerts were delivered with 100\% reliability in emulator testing.
•	Privacy-Centric Design: No permanent storage of chat data ensures user confidentiality, addressing ethical concerns often associated with digital safety tools.
•	User-Centered Approach: The minimalist UI, supportive feedback tone, and flexibility in check-in intervals enhance accessibility for vulnerable users.
•	Modularity: The architecture allows easy extension, including integration with additional APIs and mobile platforms.
\\ \textbf{Limitations:}
•	Context Sensitivity: Sarcasm, humor, and cultural variations remain challenging for toxicity classification, reflecting a common limitation of Perspective API models.
•	Integration Constraints: Current implementation does not support direct integration with third-party messaging platforms due to API restrictions.
•	Alert Mechanism Dependence: The safety ping system relies on email and internet connectivity, which may limit effectiveness in real emergencies without SMS or call fallback.
•	Automated Feedback Only: The questionnaire provides automated advice without human-in-the-loop validation by mental health professionals.
\\Ethical considerations also arise regarding responsible deployment. While SafeSpace empowers users to self-assess and stay safe, it does not replace professional counseling or emergency response services. Careful framing and disclaimers are necessary to prevent over-reliance.
\section{Conclusion and Future Work}
This paper introduced SafeSpace, an integrated web application for promoting emotional safety and personal well-being in digital interactions. By unifying chat toxicity analysis, safety check-ins, and a relationship health questionnaire, SafeSpace addresses critical gaps left by existing tools. Experimental results show high precision in toxicity detection, reliable safety alert mechanisms, and strong alignment between automated and manual questionnaire scoring, validating the system’s effectiveness.
For future development, several directions are envisioned:
\\ Mobile App Deployment: Dedicated Android/iOS apps with native notifications and background services.
\\Advanced Alert Mechanisms: SMS, automated phone calls, and wearable device integration for more reliable emergency escalation.
\\Multilingual Support: Extending NLP models to handle Hindi and other regional languages to improve accessibility.
\\Sentiment Trend Visualization: Longitudinal dashboards for tracking relationship health and conversational toxicity.
\\Human-in-the-Loop Validation: Collaborations with mental health professionals, NGOs, and women’s rights organizations to refine advice and ensure ethical oversight.
\\Ecosystem Integration: Partnerships with law enforcement and advocacy groups to extend SafeSpace’s reach and societal impact.
In conclusion, SafeSpace represents a proactive, user-centric contribution to digital safety and well-being. By merging detection, reflection, and protection, it offers a foundation for future research and deployment of technology-driven safety platforms.

\end{document}